\documentstyle[12pt]{article}
\begin{document} 
\thispagestyle{empty} 

\vspace*{4cm}

\begin{center}
{\large{\bf
Spin-field Interaction Effects and Loop Dynamics\\
in AdS/CFT Duality 
\\}}
\vspace{1.8cm} 
{\large A.~I.~Karanikas and C.~N.~Ktorides}\\ 
\smallskip 
{\it University of Athens, Physics Department\\
Nuclear \& Particle Physics Section\\
Panepistimiopolis, Ilisia GR 157--71, Athens, Greece}\\
\vspace{1cm}

\end{center}
\vspace{0.5cm}

\begin{abstract}

The spin-field interaction is considered, in the context of the gauge fields/string correspondence,
in the large 't Hooft coupling limit. The latter can be viewed as a WKB-type approximation 
to the AdS/CFT duality conjecture. Basic theoretical objects entering the present study are (a) the 
Wilson loop functional, on the gauge field side and (b) the sigma model action for the string propagating 
in AdS$_5$. Spin effects are introduced in a worldline setting, via the spin factor for a particle
entity propagating on a Wilson loop contour. The  computational tools employed
for conducting the relevant analysis, follow the methodological
guidelines introduced in two papers by Polyakov and Rychkov. The main result is expressed in terms 
of the modification of the spin factor brought about by dynamical effects, both perturbative and 
non-perturbative, according to AdS/CFT in the considered limit.

\end{abstract}


\newpage

{\bf 1. Introduction}

The connection between gauge field theories and strings has been posed, as a fundamental problem 
in theoretical physics, over two and a half decades ago by Polyakov [1]. On the field theoretical
side and in the context of 't Hooft's [2] $\lambda\equiv g_{YM}^2 N \gg 1$ 
limit, it was proposed in [3] that the
quantity upon which such a relation can be pursued is the Wilson loop functional [4]
$W[C]={1\over N}\langle Tr P\exp\oint\limits_CA_\mu dx_\mu\rangle$, whose (closed) contour
should provide a `base' on which the two ends of an (open) string, propagating in five dimensions, are 
to be attached.
The working assumption for quantifying this proposal is that, in the large $\lambda$ limit, the Wilson loop functional 
is expected to behave as
\begin{equation}
W[C]\propto e^{-\sqrt{\lambda}A_{\min}(C)},
\end{equation}
where $A_{\min}$ is the minimal area swept by the string and is bounded by the 
contour $C$. This statement constitutes a zeroth, WKB-type,
approximation to the problem. 


As is well known, in a virtual simultaneity with [3], the AdS/CFT 
conjecture [5], followed by a number
of key papers, most notably [6,7], which further elucidated its content, placed the gauge field/string duality
issue on very concrete grounds, {\it albeit} ones that favor conformally symmetric gauge 
field theories (in particular the $\cal{N}$=4 supersymmetric YM system). Within this context,
direct studies addressing themselves to the calculation of expectation values of a Wilson 
loop operator whose contour is traversed by heavy quarks, were first conducted by Maldacena in [8], followed by a
more extensive investigation in [9], as well as by Rey and Yee [10].
In these approaches, the relevant Wilson loop functional takes the 
form $W[C]={1\over N}\left < Tr P\exp\oint\limits_C(A_\mu dx_\mu+\Phi_i\,dy_i)\right>\,i=1,\cdot\cdot\cdot, 6$, 
where the $\Phi_i$ comprise a set of massive Higgs scalars, simulating heavy quarks,
in the adjoint representation of the SU(N) group.
Such considerations have direct relevance to studies of the static potential problem in QCD, {\it albeit}
in its $\cal{N}$=4 supersymmetric version.  

On the other hand and taking as point of reference work of Polchinski and Strassler [11], the
AdS/CFT framework can facilitate analyses which pertain to contributions, both perturbative and 
non-perturbative, associated with {\it dynamical}, hard scattering QCD processes.  
To the extent, then, that a Wilson loop configuration could enter a given description
referring to a {\it dynamical} process, it becomes of interest to study the effects
on the corresponding functional away from the heavy quark limit. 

Moving, now, away from the heavy quark limit inevitably brings into play the spin-field interaction term. It
is this particular issue that the present work intends to address, always in a context wherein Eq. (1) is assumed 
to be valid, {\it i.e.} in the limit of very large $\lambda$. The way by which we propose to attack the problem is the
following. First, we consider a particular casting, one we happen to be more familiar with [12], pertaining to the
worldline description of the propagation, in spacetime, 
of a (matter) particle entity which åenters a generic gauge field theoretical system. The specific feature
of this casting is that employs a quantity 
known as {\it spin factor} [13] which accounts for the particle's spin $j$. 
The relevant expression furnishes the probability amplitude 
associated with the propagation of the particle mode along a Wilson contour. The latter will be
placed in 4-dimensional space -to be viewed as the boundary of AdS$_5$- with the enclosed surface  bulging,
in general, into AdS$_5$. It will not be required for the contour to be everywhere differentiable, 
simply piecewise continuous. On the other hand, we shall not consider either loop self intersections or
retracings. The basic tool at our disposal for bringing spin effects into play is the area derivative operator [14],
which facilitates the emergence of the spin factor through the employment of a second order formalism procedure. 
This task will be presented in section 2.

In section 3, we shall direct our considerations towards the string side of the story. In particular,
we shall deal with the string action functional which furnishes the
area element associated with the  surface
swept by the `chromoelectric' string propagating in the, curved, AdS$_5$ space. The minimization of the
generated area, which is bounded by a Wilson loop, will be
discussed within the framework of the analysis devised in
two remarkable papers by Polyakov and Rychkov [15, 16]. The systematic,
variational procedure developed in these works for describing the minimum area ($A_{\min}$),
leads to the emergence of the all-important $\vec{g}(\sigma)$-function
which, in principle, carries all the dynamics (perturbative and nonperturbative) of the system. For
our particular purposes, this setting becomes quintessential in connecting the worldline formalism, that was
displayed in the previous section, with the string description. The basic
computational guidelines pertaining to our investigation will be exhibited in this section; essentially, they concern
the action of the area derivative operator on $A_{\min}$. The relevant discussion will be kept at
a general, `semiheuristic', level, leaving the display of all the technical details to an Appendix.

Our central result is exhibited in section 4, where the effect on the loop functional,  
induced by the spin-field interaction dynamics,  is formulated in terms of the $\vec{g}$-function.    
Some general suggestions pertaining to applications 
of this result are made in the concluding section. 

 \vspace*{1cm}

{\bf 2. The spin factor in the  worldline formalism}

Consider a particle entity of a given spin $j$ and with finite mass propagating on a closed worldline 
contour $\vec{c}(\sigma )$ (in Euclidean space) while interacting 
with a dynamical field $\vec{A}$. 
Following Ref. [12], the basic quantum mechanical quantity associated with the process can be written in the
form (all indices suppressed) 
\begin{equation}
K(T)= {\rm Tr}\int\limits_{c(0)=c(T)} {\cal D}c(\sigma )\exp\left(-{1\over 4}\int_0^Td\sigma\,\vec{c}'^2(\sigma )\right)
\left\langle P\exp \left(i\int_0^Td\vec{c}\cdot\vec{A}+
\int_0^Td\sigma \,F\cdot J\right)\right\rangle _A.
\end{equation}
This expression determines the probability amplitude for the system to evolve from $\vec{c}(0)$ to $\vec{c}(T)$.
The matrices $J_{\mu\nu}$ stand for the 
Lorentz generators\footnote{The term `Lorentz' is used by abuse 
of language, given that we are working in  the Euclidean formalism.} corresponding to the spin of the 
propagating particle entity, so the last term represents the spin-field interaction.

The above equation can be recast into the form
\begin{eqnarray}
&&K(T)= {\rm Tr}\int\limits_{c(0)=c(T)} {\cal D}c(\sigma )\exp\left(-{1\over 4}\int_0^Td\sigma \,
\vec{c}\,'^2(\sigma )\right)P\exp\left({i\over 2}
\int_0^Td\sigma \,J\cdot\frac{\delta}{\delta s}\right)\nonumber\\&& \quad\quad \times
\left\langle P\exp \left(i\int_0^Td\vec{c}(\sigma)\cdot\vec{A}\right)\right\rangle _A,
\end{eqnarray}
where
\begin{equation}
\frac{\delta}{\delta s_{\mu\nu}(\sigma)}
=\lim\limits_{\eta\rightarrow 0}\int\limits_{-\eta}^{\eta}dh\,h
\frac{\delta^2} {\delta c_\mu\left(\sigma+{h\over 2}\right){\delta c_\nu\left(\sigma-{h\over 2}\right)}}
\end{equation}
defines the (regularized) expression for the area derivative operator [14].

Strictly speaking, expression (3) has well defined meaning only for smooth loops. On the other hand, when
expressions like (2) are used to describe physically interesting, scattering processes the contour is forced 
to pass through points $x_i$ where momentum is imparted by an external agent (field). Such a
situation is realized by inserting a chain of delta functions $\delta[\vec{c}(\sigma_i)-\vec{x}_i]$ in the integrand
which produce a loop with cusps. Accordingly, the area derivative operator entering (3) must be understood
piecewise, {\it i.e.},
\begin{equation}
P\exp\left({i\over 2}\int_0^Td\sigma \,J\cdot\frac{\delta}{\delta s}\right)\rightarrow
\cdot\cdot\cdot P\exp\left({i\over 2}\int_{s_1}^{s_2}d\sigma \,J\cdot\frac{\delta}{\delta s}\right) 
P\exp\left({i\over 2}\int_0^{s_1}d\sigma \,J\cdot\frac{\delta}{\delta s}\right).
\end{equation}

An integration by parts can now be performed to reformulate the amplitude (3) as follows 
\begin{eqnarray}
&&K(T)= {\rm Tr}\int\limits_{c(0)=c(T)} {\cal D}c(\sigma)\exp\left(-{1\over 4}\int_0^Td\sigma\,
\vec{c}\,'^2(\sigma)\right)
P\exp\left({i\over 2}\int_0^Td\sigma \,J\cdot\omega(c)\right) \nonumber\\&& \quad\quad \times
\left\langle P\exp \left(i\int_0^Td\vec{c}\cdot\vec{A}\right)\right\rangle _A,
\end{eqnarray}
where
\begin{eqnarray}
\int_0^Td\sigma\,\omega_{\mu\nu}(c)&=&\lim\limits_{\eta\rightarrow 0}{1\over 2}\int_{-\eta}^\eta dh\,h
\int_0^Td\sigma \,c''_\mu\left(
\sigma -{h\over 2}\right)c''_\nu\left(\sigma +{h\over 2}\right)\nonumber\\
&=&{1\over 4}\lim\limits_{\eta\rightarrow 0}\int_{-\eta}^\eta dh\int_0^Td\sigma _1\int_0^Td\sigma_2\,
c'_{[\mu}(\sigma_2)c''_{\nu]}(\sigma_1)
\delta(\sigma _2-\sigma_1-h).
\end{eqnarray}  

The quantity $\omega_{\mu\nu}(c)$ defines the {\it spin factor} [12] associated with the
particle mode propagating on the closed contour and has a geometrical/topological content. 
Thus, for example, in two dimensions it
serves to distinguish free bosons from free fermions by the fact that a member 
of the latter species carries a factor $(-1)^\nu$ when traversing a closed worldline
contour $\nu+1$ times. Our goal is to identify {\it dynamical} consequences associated with its presence 
which, by definition, are attributable to the spin of a particle mode propagating on the Wilson loop contour. 
As already mentioned, our attention will be restricted, throughout, to {\it single} traversals 
of, non self-intersecting, loops. 

A general observation to make at this point is that Eq (7) leads to 
a vanishing result for the spin factor if, in the absence of the Wilson loop, the worldline contour is everywhere 
differentiable, since the presence of a factor $\delta'(h)$ is required for the opposite to be the case.
Non-trivial, dynamically induced, effects attributed to spin thereby {\it demand} the presence of Wilson loop 
configurations which include points of interaction of the propagating entity with the dynamical
field. In a perturbative treatment of such a situation one expands the Wilson loop in a power series which
produces the familiar vertices. Explicitly, the presence of the spin factor gives rise to correlators
of the form 
\begin{equation}
\lim\limits_{\eta\rightarrow 0}\int_{-\eta}^\eta dh\,h\int_0^Td\sigma\left< c''_\mu\left(
\sigma -{h\over 2}\right)c''_\nu\left(\sigma +{h\over 2}\right)c'_k(\sigma_1)c'_\lambda(\sigma_2)\right>_C
\sim\delta(\sigma_2-\sigma_1)(\delta_{\mu\kappa}\delta_{\nu\lambda}-\delta_{\mu\lambda}\delta_{\nu\kappa}),
\end{equation}
where $\langle\cdot\cdot\cdot\rangle_C$ signifies averaging over paths.

It is evident from the above analysis that the spin factor incorporates, in a geometrical manner, the 
spin-field dynamics. The challenge, now, can be described as follows: Determine the expression for the
minimal area on the string side and, 
upon doing that,  use the operation in (3) to assess the nonperturbative, dynamical impact on the spin factor.

\vspace*{1cm}

{\bf 3. The area derivative of the Wilson loop functional}

In Refs [15,16] a mathematical machinery was developed for the purpose of studying loop dynamics
in the framework  of the AdS/CFT correspondence (in the WKB approximation). We shall adopt the strategy 
introduced in these works
with the eventual aim being the determination of the action of the area derivative, as given by (4), on a, piecewise 
continuous, Wilson loop functional -as demanded by Eq. (3). The dynamics
of the chromo-electric flux lines is described, according to [3,8,16],
by a relativistic string propagating in a five-dimensional (AdS) curved background. The relevant 
action functional (Euclidean formalism adopted throughout our analysis) is given by [15]
\begin{eqnarray}
S[\vec{x}(\xi),y(\xi)]&=&{1\over 2}\sqrt{\lambda}\int_D G_{MN}(x(\xi))\partial_ax^M(\xi)\partial_a x^N(\xi)\nonumber\\
&=&
{1\over 2}\sqrt{\lambda}\int_D 
\frac{d^2\xi}{y^2(\xi)}
[(\partial_a \vec{x}(\xi))^2+(\partial_a y(\xi))^2],
\end{eqnarray} 
where $x^M=(y,\vec{x})=(y,x^\mu),\, M,N=0,1,\cdot\cdot\cdot,4;\,\mu=1,\cdot\cdot\cdot,4$, with
the $y$-coordinate taking a zero value at the boundary and growing toward infinity as
one moves deeper into the interior of the AdS$_5$ space. 

The above functional is to be minimized under the boundary conditions $\vec{x}|_{\partial D}=\vec{c}(\alpha(\sigma))$   
and $y|_{\partial D}=0$, with the parametrization chosen so that  
\begin{equation}
A_{\min}[c(\sigma)]=\min\limits_{\{\alpha(\sigma)\}}\min\limits_{\{\vec{x},y\}}S[\vec{x}(\xi),y(\xi)].
\end{equation}
The functional $A_{\min}$ is invariant under reparametrizations of the boundary, a property that can be easily 
deduced  from the minimization condition (10):
\begin{equation}
c'_\mu(\sigma)\frac{\delta A_{\min}}{\delta c_\mu(\sigma)}=0.
\end{equation}

Following Refs [15,16], we adopt the static gauge $y(t,\sigma)=t$ and place the 
loop on the boundary of the AdS space, i.e. set $t= 0$. One, accordingly, writes
\begin{equation}
\vec{x}(t,\sigma)=\vec{c}(\sigma) +{1\over 2}\vec{f}(\sigma)t^2+{1\over 3}\vec{g}(\sigma)t^3+\cdot\cdot\cdot
\end{equation}
where, for now, the curve $\vec{c}(\sigma)$ is assumed to be 
everywhere differentiable. If there are cusps 
on the loop contour ({\it i.e.}, discontinuities in the first derivative) the above expansion must be 
understood piecewise. Surface minimization eliminates the linear term in the expansion 
and determines its next coefficient:
\begin{equation}
\vec{f}=\vec{c}'^2\frac{d}{d\sigma}\frac{\vec{c'}}{\vec{c'}^2}.
\end{equation}
The coefficient  $\vec{g}(\sigma)$ is, at this point, unspecified.
Employment of the Virasoro constraints leads to
\begin{equation}
\vec{c'}\cdot\vec{g}=0.
\end{equation}
It turns out that the latter relation simply expresses the reparametrization invariance 
of the minimal area (10) and, hence, the quantity
$\vec{g}(\sigma )$, to be referred to as $\vec{g}$-function from hereon, remains undetermined. More illuminating,
for our purposes, is an interim result through which (14) is derived and reads as follows 
\begin{equation}
\frac{\delta A_{\min}}{\delta\vec{c}(\sigma)}=-\sqrt{\vec{c'}^2}\vec{g}(\sigma).
\end{equation}
The above relation underlines the dynamical significance of the $\vec{g}$-function: It provides
a measure of the change of $A_{\min}$ when the Wilson loop contour is altered as a result of some
interaction which reshapes its geometrical profile. 

Consider now the action of the area derivative on the Wilson loop functional:
\begin{equation}
\frac{\delta}{\delta s_{\mu\nu}(s)}W[C]=\lim\limits_{\eta\rightarrow 0}\int\limits_{-\eta}^\eta dh\,h
\left[-\sqrt{\lambda}
\frac{\delta^2 A_{\min}} {\delta c_\mu\left(\sigma+{h\over 2}\right){\delta c_\nu\left(\sigma-{h\over 2}\right)}}
+\lambda\frac{\delta A_{\min}} {\delta c_\mu\left(\sigma+{h\over 2}\right)}
\,\frac{\delta A_{\min}} {\delta c_\nu\left(\sigma-{h\over 2}\right)}\right]W[C].
\end{equation}
As it is known [17], the area derivative is a well defined operation only for smooth contours, i.e.
everywhere differentiable. In such a case the last term in the above equation gives zero contribution. If the loop under 
consideration has cusps (as happens in the framework of non-trivial applications 
of the worldline formalism) the operation must be understood piecewise [18].

In order to facilitate our considerations we follow Ref(s) [15, 16] by choosing the
coordinate $\sigma$ on the minimal surface such that
\[
\vec{c}'^2(\sigma)=1,\quad \dot{\vec{x}}(t,\sigma)\cdot \vec{c}'(\sigma)=0.
\]
We also introduce an orthonormal basis, which adjusts itself
along the tangential ($\vec{t}$) and normal ($\vec{n}^a\, ,a=1,\cdot\cdot\cdot,D-1$) directions defined by the
contour, as follows
\begin{eqnarray}
&&\{\vec{t}, \vec{n}^a\},\,a=1,\cdot\cdot\cdot,D-1\nonumber\\ 
&& \vec{t}=\frac{\vec{c}\,'}{\sqrt{\vec{c}\,^2}},\quad\vec{n}^a\cdot\vec{t}=0,\quad \vec{n}^a
\cdot\vec{n}^b=\delta^{ab}.
\end{eqnarray}
We now write
\begin{equation}
\frac{\delta}{\delta c_\mu}=n_\mu^a\left(\vec{n}^a\cdot\frac{\delta}{\delta \vec{c}}\right)
+t_\mu\left(\vec{t}\cdot\frac{\delta}{\delta \vec{c}}\right)\equiv n_\mu^a\frac{\delta}{\delta\vec{n}^a}
+t_\mu\frac{\delta}{\delta \vec{t}} 
\end{equation}
and upon using relations (14) and (15), as well as setting $s=\sigma+h/2$ and $s'=\sigma-h/2$,  we determine
\begin{equation} 
\frac{\delta^2 A_{\min}} {\delta c_\mu(s)\delta c_\nu(s')}=-\frac{\delta g^a(s)}{\delta\vec{n}^b(s')}
n_\mu^a(s)n_\nu^b(s')+R_{\mu\nu}(s,s')\delta'(s-s'),
\end{equation}
where
\begin{equation}
R_{\mu\nu}(s,s')=2\vec{g}(s)\cdot\vec{n}^a(s')t_\mu(s) n^a_\nu(s') + \vec{g}(s)\cdot\vec{t}(s')t_\mu(s) t_\nu(s')
- \vec{t}(s)\cdot\vec{n}^a(s')g_\mu(s)n_\nu^a(s').
\end{equation}
Given the defining expression for the area derivative, cf. Eq (4), 
one immediately realizes that only terms $\sim \delta '(s-s')$ in an
antisymmetric combination $R_{[\mu\nu]}$ will give non-zero contributions to the area derivative. 
It, thus, becomes obvious that the last term in Eq (19) produces the result
\begin{equation}
R_{[\mu\nu]}(\sigma,\sigma) = t_{[\mu}(\sigma) g_{\nu]}(\sigma).
\end{equation}

Turning our attention to the first term on the rhs of (20) we note that non-vanishing contributions 
should have the form  
\begin{equation}
(r^aq^b-r^bq^a)n_\mu^an_\nu^b\delta'(s-s'),
\end{equation}
where $r^a=\vec{n}^a\cdot\vec{r}$ and $q^a=\vec{n}^a\cdot\vec{q}$. These functions must be determined from the 
coefficients of the expansion (12); otherwise the above contribution would be contour independent, having no impact 
on a calculation associated with non-trivial dynamics. In conclusion, a simple qualitative analysis, based
on the scale invariance of $A_{\min}$, indicates that a contribution of the type (22) does not exist. 
This qualitative observation can be further substantiated through a straigtforward argument based on
dimensional grounds.
Indeed, from Eq. (12) it can be observed that under a change of scale of the form
$\vec{c}\rightarrow \lambda\vec{c},\,(t,\sigma)\rightarrow (\lambda t,\lambda \sigma)$ one has
\[
\vec{c}'\rightarrow \vec{c}',\quad \vec{f}\rightarrow{1\over\lambda} \vec{f},\quad 
\vec{g}\rightarrow {1\over \lambda^2}\vec{g},\cdot\cdot\cdot.
\]
On the other hand the area derivative, being of second order, should scale $\sim {1\over \lambda^2}$. In turn,
this means that one of the quantities $\vec{r}$ or $\vec{q}$, which must arise through transverse variations
of $\vec{g}$, should be aligned with the tangential vector $\vec{t}$, which , by definition, has zero transverse 
componenents. An explicit verification of the result prompted by the preceding,
 heuristic arguments, is presented in the Appendix.
In the course of that computation the following relation is obtained (the {\it tilde} denotes a Fourier transformed
quantity, to be defined below)
\begin{eqnarray}
\delta\tilde{g}^a(p)&=&\left[ |p|^3\delta^{ab} -|p|(\vec{f}^2\delta^{ab}-3f^af^b)\right]n_\mu ^b\delta\tilde{c}_\mu (p) 
\nonumber\\ & &-\left[\vec{f}\cdot\vec{g}\delta^{ab}-{3\over 2}(f^ag^b+f^bg^a)-{25\over 12}(f^ag^b-f^bg^a)\right]
n_\mu^b\delta \tilde{c}_\mu(p)+\cdot\cdot\cdot,
\end{eqnarray}
where the $p$ variable enters through a Fourier transform specified by
\begin{equation}
F(s)=F(s'+h)=\int\frac{dp}{2\pi}e^{iph}\tilde{F}(s',p).
\end{equation}  
Since $|h|<\eta\rightarrow 0$, what we have examined is the variation of $\tilde{g}^a$ for $|p|\rightarrow\infty$.
The dots in (23), accordingly, represent terms that vanish as $|p|^{-1}$. It also follows 
from (24) that all the functions on the rhs of (23) are calculated at $s'$.
We thereby deduce that 
\begin{eqnarray}
\frac{\delta\tilde{g}^a(s)} {\delta\vec{n}^b(s')}
&=&\int \frac{dp}{2\pi}
\left[
|p|^3\delta^{ab} -|p|(\vec{f}^2\delta^{ab}-3f^af^b)
\right]e^{iph} 
\nonumber\\ & &-
\left[\vec{f}\cdot\vec{g}\delta^{ab}-{3\over 2}(f^ag^b+f^bg^a)-{25\over 12}(f^ag^b-f^bg^a)
\right]
\delta(h)+{\cal O}(h).
\end{eqnarray}

Referring to the formula for the area derivative, 
we immediately surmise that the first term on the rhs  of Eq. (20) gives null 
contribution since the antisymmetric term in (25) is
proportional to $\delta(s-s')$, 
and {\it not} $\delta'(s-s')$. We have, therefore, determined that
\begin{equation} 
\lim\limits_{\eta\rightarrow 0}\int\limits_{-\eta}^\eta dh\,h
\frac{\delta^2 A_{\min}} {\delta c_\mu\left(\sigma+{h\over 2}\right){\delta c_\nu\left(\sigma-{h\over 2}\right)}}
=t_{[\mu}\sigma)g_{\nu ]}(\sigma)
\end{equation}

Ending this section we find it useful to apply the above result for the purpose of verifying the 
Makkenko-Migdal equation [19], see Refs. [20] for review expositions, 
for a {\it differentiable}, non-selfintersecting Wilson loop which is traversed only once, namely 
\begin{equation}
\tilde{\Delta}W[C]\approx 0,
\end{equation}
where the symbol $\approx$ means that the finite part on the rhs is zero
and the MM loop operator is defined [20] as
\begin{equation}
\tilde{\Delta}=\oint dc_\nu\partial_\mu^{c(\sigma)} \frac{\delta}{\delta s_{\mu\nu}(\sigma)}=
\lim\limits_{\eta\rightarrow 0}\lim\limits_{\eta'\rightarrow 0}\int d\sigma\,c_\nu(\sigma)
\int\limits_{\sigma -\eta}^{\sigma +\eta} d\sigma'\frac{\delta}{\delta c_\mu(\sigma')}\int\limits_{-\eta'}^{\eta'}
dh\,h\frac{\delta^2}{\delta c_\mu(\sigma +h)\delta c_\nu(\sigma)}.
\end{equation}

It can, now, be easily determined that
\begin{equation}
\tilde{\Delta}A_{\min}=2\lim\limits_{\eta\rightarrow 0}\int d\sigma\,c_n (\sigma)\int\limits_{\sigma-\eta}^{\sigma+\eta}
d\sigma'\frac{\delta}{\delta c_\mu(\sigma')}[t_{\nu}(\sigma)g_\mu(\sigma)]=2\lim\limits_{\eta\rightarrow 0}
\int d\sigma\int\limits_{\sigma-\eta}^{\sigma+\eta}d\sigma'
\frac{\delta g_\mu(\sigma)}{\delta c_\mu(\sigma')}.
\end{equation}
But
\begin{eqnarray}
&&\frac{\delta g^a(\sigma)}{\delta 
\vec{n}^b(\sigma')}n^a(\sigma)\cdot n^b(\sigma')=-(D-4)\vec{f}\cdot\vec{g}\delta(\sigma-\sigma')+\nonumber\\
&&+\left[\frac{3!}{\pi}\frac{\delta^{ab}}{(\sigma-\sigma')^4}+{1\over \pi}\frac{1}{(\sigma-\sigma')^2}
(\vec{f}^2\delta^{ab}-3f^af^b)\right]
\vec{n}^a(\sigma)\cdot\vec{n}^b(\sigma')+{\cal O}(\sigma-\sigma')
\end{eqnarray}
and so
\begin{equation}
\tilde{\Delta}A_{\min}=2\lim\limits_{\eta\rightarrow 0}\int d\sigma\int\limits_{\sigma-\eta}^ {\sigma+\eta}
d\sigma'
\frac{\delta g^a(\sigma)}{\delta \vec{n}^b(\sigma')}
\vec{n}^a(\sigma)\cdot\vec{n}^b(\sigma')\approx 0.
\end{equation}

\vspace*{1cm}

{\bf 4. The spin factor contribution and the role of the $\vec{g}$-function}

Using the fact that the area derivative operator obeys the Leibnitz rule, we  
obtain, after appplying it twice on the Wilson loop functional
and using Eq. (26),
\begin{equation}
\frac{\delta}{\delta s_{\kappa\lambda}(\sigma')}\frac{\delta}{\delta s_{\mu\nu}(\sigma)}W
=-\sqrt{\lambda}t_{[\mu}(\sigma)g_{\nu ]}(\sigma)\frac{\delta}{\delta s_{\kappa\lambda}(\sigma')}W=\lambda  
t_{[\mu}(\sigma)g_{\nu]}(\sigma)t_{[\kappa}(\sigma')g_{\lambda]}(\sigma')W.
\end{equation}

Its action, as is evident from Eq (3), appears in an exponentiated form: 
\begin{eqnarray}
P\exp\left({i\over 2}\int d\sigma\, J\cdot\frac{\delta}{\delta s}\right)W[C]&\propto& 
P\exp\left({i\over 2}\int d\sigma\, J\cdot\frac{\delta}{\delta s}\right)e^{-\sqrt{\lambda}
A_{\min}[C]}\nonumber\\&&
\propto P\exp\left[{i\over 2}\sqrt{\lambda}\int d\sigma\, J_{\mu\nu}t_{[\mu}(\sigma)g_{\nu]}(\sigma)\right]e^{-\sqrt{\lambda}
A_{\min}[C]}.
\end{eqnarray}
Once again, the operation of the area derivative must be understood piecewise if the Wilson
loop configuration has cusps. In such a case the spin factor contribution factorizes
into pieces, each one of which represents `spin factor' expressions associated with the corresponding smooth 
component [14] of the piecewise connected contour.

Further progress can be made if one imposes, in the $\sqrt{\lambda}\gg 1$ limit at least, the
Bianchi identity at every point at which the Wilson contour is smooth (differentiable). The latter reads
\begin{equation}
\epsilon_{\kappa\lambda\mu\nu}\partial^{c(\sigma)}_\lambda 
\frac{\delta}{\delta s_{\mu\nu}(\sigma)}W=\lim\limits_{\eta\rightarrow 0}\epsilon_{\kappa\lambda\mu\nu}
\int\limits_{\sigma-\eta}^{\sigma+\eta}d\sigma'
\frac{\delta}{\delta c_\lambda(\sigma')}
\frac{\delta}{\delta s_{\mu\nu}(\sigma)}W=0.
\end{equation}

Substitution of the result (26) for the area derivative into the Bianchi identity, leads to
\begin{equation}
\epsilon_{\kappa\lambda\mu\nu}\partial_\lambda^{c(\sigma)}\frac{\delta}
{\delta s_{\mu\nu}(\sigma)}W=\lim\limits_{\eta\rightarrow 0}
\int\limits_{\sigma-\eta}^{\sigma+\eta}d\sigma'\frac{\delta}{\delta c_\lambda(\sigma')}
\left[t_{[\mu}(\sigma)g_{\nu]}(\sigma))\right]=0.
\end{equation}

Referring, now, to Eq. (20), one finds
\begin{eqnarray}
&&t_\mu(\sigma)\frac{\delta g_\nu(\sigma)}{\delta c_\lambda(\sigma')}-
(\mu\leftrightarrow\nu)=\frac{\delta g^a(\sigma)}{\delta\vec{n}^b(\sigma')}n^b_\lambda(\sigma)
t_{[\mu}(\sigma)n_{\nu]}^a(\sigma)\nonumber\\ &&
\quad\quad -\delta'(\sigma-\sigma')\vec{t}(\sigma)\cdot \vec{n}^a(\sigma')n_\lambda^a(\sigma)
t_{[\mu}(\sigma)g_{\nu]}(\sigma),
\end{eqnarray}
which finally gives
\begin{equation}
\epsilon_{\kappa\lambda\mu\nu}\lim\limits_{\eta\rightarrow 0}\int\limits_{\sigma-\eta}^{\sigma+\eta}d\sigma'
\frac{\delta g^a(\sigma)}{\delta\vec{n}^b(\sigma')}n_\lambda^b(\sigma')
t_{[\mu}(\sigma)n_{\nu]}^a(\sigma)=\epsilon_{\kappa\lambda\mu\nu}c''_\lambda(\sigma)
t_{[\mu}(\sigma)g_{\nu]}(\sigma).
\end{equation}

Substituting into the above equation the result expressed by Eq (25), one  concludes that
\begin{equation}
\epsilon_{\kappa\lambda\mu\nu}c''_\lambda(\sigma)t_{[\mu}(\sigma)g_{\nu]}(\sigma)=0.
\end{equation}
Given, now, that $\vec{g}\cdot\vec{c}'=0,\,\vec{c}''\cdot \vec{c}'=0$, we surmise that vectors $\vec{g}$ and $\vec{c}''$
are parallel to each other. Accordingly, we write $\vec{g}(\sigma)=\phi(\sigma)\vec{c}''(\sigma)$,
with $\phi=\frac{|\vec{g}|}{|\vec{c}''|}$. In turn,
this leads to the deduction that, for a smooth Wilson contour, the following holds true
\begin{equation}
P\exp\left(i{i\over 2}\int d\sigma J_{\mu\nu}\frac{\delta}{\delta s_{\mu\nu}}\right)W[C]\propto 
P\exp\left({i\over 2}\int d\sigma \phi(\sigma)c'_{[\mu}(\sigma)c''_{\nu]}(\sigma)J_{\mu\nu}\right)W[C].
\end{equation}
The above relation constitutes the central result of this paper. It exhibits a `deformed' spin factor
expression which incorporates dynamical effects induced via the AdS/CFT duality conjecture, in the
$\sqrt{\lambda}\gg 1$ limit. Comments and/or speculations surrounding this result will be presented 
in the concluding section which follows. 

\vspace*{1cm}

{\bf 5. Assessments and concluding remarks}

To initiate an evaluation, from a physics point of view, of implications of the analysis conducted in 
this paper, let us start by making some general comments in reference to the
issue of perturbative vs. nonperturbative 
aspects of QCD, {\it as a quantum field theory}. For our starting point, we adopt the universally accepted
belief that lattice QCD consitutes the most effective 
approach for  the study of nonperturbative phenomena in the theory. As other, non-lattice, examples\footnote{By 
no means does this exhaust the list 
of all, relevant, theoretical proposals.} of, credible, attempts for non-perturbative, field theoretical 
investigations of the theory one could mention: (a) The 
loop equation approach [17,20] and (b) the Stochastic Vacuum Model [21]. In all these cases, the Wilson 
loop, which enters either directly or via the 
non-abelian Stoke's theorem, constitutes a fundamental element of the corresponding 
formulations. Perturbation theory, on the 
other hand, bases its description strictly on locality premises.

With the above in place, consider a dynamical process involving fundamental particle 
entities, {\it e.g.} quarks, whose propagation is described 
in terms of worldline contours, in line, for example, with Eq. (3). Such a description mode adopts, just 
like string theory does, 
first quantization methods, as opposed to field theoretical formulations 
which adhere to a second quantization methodology. Suppose, now, that  the worldline description 
of a given (QCD) process of interest involves closed worldline paths. Then, the interactions 
of the propagating particle entity with gauge fields, generate
Wilson contours. If, now, (local) interaction(s) with some external agent(s) take place, then the contour 
will be deformed through the formation of cusps, i.e. points at which a four-momentum is imparted.
Our working premise is that perturbative contributions to the process correspond to, local gluon exchanges, 
as well as their emission and/or absorption, with point of reference a corresponding cusp vertex.
Non-perturbative, dynamical effects, on the other hand, should be associated with area deformations. 

The preceeding, intuitive, comments are very general and pertain to QCD as a quantum field theory,
in a wider sense. The AdS/CFT setting, on the other hand, corresponds to a theory which is characterized as 
`holographic QCD'. Setting aside the issue regarding the precise connection between the quantum 
field theoretical and the holographic version of QCD, equivalently, the precise 
connection between gauge fields and strings, let us adopt as working hypothesis that
the WKB-type approximation adopted in this study constitutes a zeroth approximation to QCD whose
basic merit is that it contains both perturbative and non-perturbative contributions
to dynamical physical processes. From such a prespetive, the primary issue of relevance is to gain a 
concrete, {\it quantitative} perspective on the $\vec{g}$-function. 
So, let us assume that our main result, as expressed in general form by Eq. (33), gives the leading contribution
to a given physical process, 
which involves an integral over Wilson contours with certain characteristics. As it stands,
it tells us that the dominant, {\it piecewise continuous} contours are those for which $\vec{g}=0$ 
and, consequently, see Eqs. (33) and (39), the spin factor becomes unity. 
By itself, this occurence constitutes a consistency check with existing applications of the
worldline formalism to situations where the eikonal approximation is valid [22], {\it i.e.} when the dominant
contour is formed by straight line segments. On the other hand, a piecewise continous contour
does not have to be of a polygonal type. Distorsions, induced by interactions which 
keep the path segments smooth while changing A$_{\min}$ could very well give the dominant contribution.
In other words, dynamical, non-perturbative information 
may very well reside in solutions of equation $\vec{g}=0$. Reversing the argument, suppose one 
is in position to surmise the worldline, geometrical profile of a Wilson contour 
associated with a given dynamical process on purely physical grounds. Then, one could be in
position to {\it a priori} determine A$_{\min}$. Such scenarios {\it are}
realistic and have, in fact, been studied, perturbatively, in the worldline context,
see [22] and references therein. Non-perturbative dynamical contributions stemming from 
the WKB approximation to holographic QCD should enter as further correction terms 
to well established perturbative expressions based on resummation procedures, once some
input for the $g$-function, phenomenological or, model-dependent, theoretical is provided.

As a final note of interest we consider the
so called wavy line approximation [16], according to which the Wilson contour, in four dimensions,
is parametrized as $(\sigma,\psi_i),\,i=1,2,3$, with the $\psi_i$ small transverse deviations.
One finds, on account of reparametrization invariance and the Hamilton-Jacobi equations
for the minimal area [16], that
\[
\left.\frac{\delta A_{\min}}{\delta \psi^i}\right|_{\psi^i=0}=0.
\]
The above result gives another perspective on why for heavy quarks -as well as other situations for
which the eikonal approximation is valid- (piecewise) straight Wilson paths play the dominant 
role: The $\vec{g}$- function vanishes in this case. 
Correction terms will arise, on the other hand, only 
if transverse fluctuations are taken into account. As alluded to already, any distortion of a given path segment,
which keeps it smooth, while the $\vec{g}$-function remains zero, could have, in principle, non-trivial
significance which will be reflected in the expression for A$_{\min}$.

\newpage

\appendix
\setcounter{section}{0}
\addtocounter{section}{1}
\section*{Appendix}
\setcounter{equation}{0}
\renewcommand{\theequation}{\thesection.\arabic{equation}}

Here we shall give a proof of Eq. (23) in the text following closely the methodology of Ref [15,16].

As is obvious from Eq. (18), in order to compute the area derivative we need the normal variation 
of the $\vec{g}$-function. As a first step in this direction one defines, at every point of the surface,
a basis $\{n^a_M(t,s)\}$ of $D-1$ orthonormal vectors which satisfy the conditions 
\begin{equation} 
n_M^a (t,s)\dot x_M (t,s) = n_M^a (t,s)x'_M (t,s) = 0,
\end{equation}
where $G_{MN}n_M^a n_N^b=\delta^{ab}$ and $n_\mu^a(0,s)=n_\mu^a(s)$ are the vectors used in Eq. (19) of the text.

Under the normal variation                
\begin{equation} 
x_M (t,s) \to x_M (t,s) + \psi _M (t,s),\quad \psi _M (t,s) = \phi ^a (t,s)n_M^a (t,s)
\end{equation}
the change of the minimal surface to second order in $\phi^a$ reads
\begin{equation}
S^{(2)}  = \int {d^2 } \xi \left[ {\sqrt g (g^{\alpha \beta } \partial _\alpha  \psi ^a 
\partial _\beta  \psi ^a  + 2g^{\alpha \beta } \omega _\alpha ^{[ab]} \partial _\beta  
\psi ^a \psi ^b  + 2\psi ^a \psi ^a ) + O(t^2 \psi ^2 )} \right]
\end{equation}
where we have written $\psi^a\equiv t\phi^a$ and have introduced $g_{\alpha\beta}=G_{MN}
\partial_\alpha x_M\partial_\beta x_N$, while the, antisymmetric, quantities $\omega _\alpha ^{[ab]}$ are spin connection 
coefficients and are given by
\begin{equation}
\omega _\alpha ^{[ab]}  = \partial _\alpha  n_M^a  \cdot n_M^a 
\end{equation}  

The exact form of this result can be found in [16]. Here, all we need is the third order term in an expansion
of $\psi_M$ in powers of $t$. Taking into account that $\phi$ is regular as $t\to 0$, we have ommitted terms 
$\sim t^4$ in (A.3) which do not contribute to the normal variation of the $\vec{g}$-function.
                                                                                                                               
Using the expansion (12) one easily determines that
\begin{equation}
g_{\alpha \beta }  = {1 \over {t^2 }}\left( 
\begin {array}{ll}
  1 + \vec f^2 t^2+ 2\vec f \cdot \vec gt^3  \quad\quad\quad\quad\quad\quad{1\over 2}\vec f \cdot \vec f't^3\\ 
  \quad\quad{1\over 2}\vec{f} \cdot \vec {f}'t^3 {\rm  }\quad\quad\quad\quad 1 - {{\rm 1} \over {\rm 2}}\vec f^2 t^2  - 
{2 \over 3}\vec f \cdot \vec{g}t^3   + O(t^2 )\\
\end{array}\right)
\end{equation}
and
\begin{equation}                                                                                                       
\sqrt g  = {1 \over {t^2 }}(1 + {2 \over 3}\vec f \cdot \vec gt^3 ) + O(t^2 ).
\end{equation}                                                                                                                        
                                                                                                                                 
Now, the area derivative receives contributions from antisymmetric terms. We, therefore, have to find the behavior
of the spin connection as $t\to 0$. This cannot be done in a unique way if $D>2$. What one can do is to expand
the basis vectors $n_M^a(t,s)$ as a power series in $t$:
\begin{eqnarray}                                      
&&n_0^a (t,s)= tk_0^a (s)+{1 \over 2}t^2 l_0^a (s) + {1 \over 3}t^3 m_0^a (s) +\cdot\cdot\cdot\nonumber\\
&&\vec n_{}^a (t,s) = t\vec k_{}^a (s)+{1 \over 2}t^2 \vec l_{}^a (s) + {1 \over 3}t^3 \vec m_{}^a (s) +\cdot\cdot\cdot
\end{eqnarray}

Combining these relations with (A.1) and using the expansion (12) we can determine that                                                                                                                               
\begin{equation}
k_0^a  = f^a, \,\,l_0^a =-2(\vec k^a  \cdot \vec f + g^a ),\,\,m_0^a=  
- 3({1 \over 2}\vec l^a  \cdot \vec f + \vec k^a  \cdot \vec g + h^a )
\end{equation}
and                                                                                                                                 
\begin{equation}
\vec k^a \cdot \vec c'=0,\,\,\vec l^a  \cdot \vec c'+ f'^a  =0,\,\, 
\vec m^a  \cdot \vec c' + g'^a  + {3 \over 2}\vec k^a  \cdot \vec f = 0.
\end{equation}
                                                                                                                                 
From the orthonormality condition we find that
\begin{eqnarray}
 &&\vec k^a  \cdot \vec n^b (s) + \vec k^b  \cdot \vec n^a (s) = 0,\,\,
2k_M^a  \cdot k_M^b  + \vec l^a  \cdot \vec n^b (s) + \vec l^b  \cdot \vec n^a (s) = 0\nonumber\\
&&\quad\quad\quad\quad {3 \over 2}l_M^a  \cdot l_M^b  + \vec m^a  \cdot \vec n^b (s) + \vec m^b  \cdot \vec n^a (s) = 0.
\end{eqnarray}
                                                                                                                                 
With the above in place we return to our central objective and, to start with, assume that
\begin{equation}
\vec k^a  \cdot \vec c' = 0{\rm   } \to {\rm   }\vec k^a  = \vec 0, 
\end{equation}                     
which means that
\begin{equation}
\begin{array}{cc}
\vec l^a  \cdot \vec c' =  - f^a\\ 
\vec l^a  \cdot \vec n^b (s) + \vec l^b  \cdot \vec n^a (s) =  - 2k_0^a k_0^b  =  - 2f^a f^b.\\
\end{array} 
\end{equation}
                                                                                                                               
From these relations we conclude that
\begin{equation}
{\rm   }\vec l^a  =  - f'^a \vec c' - f^a \vec f + \Lambda ^{ab} \vec n^b (s).
\end{equation}    
with $\Lambda^{ab}$ antisymmetric, but otherwise arbitrary. The observation here is the following: 
On the one hand $\Lambda^{ab}$ enters the second order term of the expansion (A.6) and consequently contributes
to the {\it normal} variation of the $\vec{g}$-function, to the area derivative and to the spin factor. 
On the other hand, it does not depend on the functions $\vec{c}', \vec{f},\vec{g},\cdot\cdot\cdot$
which determine $A_{\min}$. This can be deduced, through scaling prooperties as follows: Under a change
of scale $\vec{c}\to\lambda\vec{c}, (t,s)\to\lambda(t,s)$, it must behave as $\Lambda\to{1\over \lambda^2}$, as
can be seen from Eq. (A.6). Taking, now, into account that 
$\vec{c}'\to\vec{c}',\,\vec{f}\to{1\over \lambda}\vec{f},
\,\vec{g}\to{1\over \lambda^2}\vec{g},\cdot\cdot\cdot$ 
and that $\vec{n}^a(s)\cdot\vec{c}'=0\to c'^a=0$,
it becomes obvious that it is impossible to find an antisymmetric combination of the coefficient functions to construct
$\Lambda^{ab}=-\Lambda^{ba}$. Thus, this arbitrary function does not depend on the contour and 
consequently can be chosen at will. We shall take it to be zero.With the same reasoning assumption (A.10) can be 
justified and so we can determine the basis vectors:
\begin{equation}
\begin{array}{cc}
n_0^a (t,s) =  - tf^a  - t^2 g^a  - t^3 (h^a  - f^a \vec f^2 ) + O(t^4 )\\
 \vec n^a(t,s)=\vec n^a (s)-{1\over 2}t^2(f^a\vec f + f'^a \vec c')-{1 \over 2}t^3 
(g^a \vec f + f^a \vec g + {2 \over 3}g'^a \vec c') + O(t^4 )\\
\end{array}
\end{equation}
For the behavior of the spin connection we also need the derivative $\vec{n}'â(s)$. What we do know about it comes 
from the orthonormality condition
\begin{equation}
\vec n^a (s) \cdot \vec c'=0{\rm   }\to {\rm   }\vec n'^a (s) \cdot \vec c'= - \vec n^a (s) \cdot \vec c''
\end{equation}
Adopting the same arguments as before we conclude from the preceeding relation that
\begin{equation}
\vec n'^a (s) =  - (\vec n^a (s) \cdot \vec c'')\vec c' =  - c''^a \vec c'
\end{equation}
In conclusion, through the above analysis we have determined that
\begin{equation}
\omega _t^{[ab]}={1 \over 2}t^2 (g^a f^b- g^b f^a )\equiv{1 \over 2}t^2 r^{ab},\quad\omega _s^{[ab]}
= {\cal O}(t^3 ).
\end{equation}
Knowing the behavior of all the terms we now return to (A.3) and demand the perturbed surface also to be minimal.
This leads to the equation 
\begin{equation}                                                                                                                              
 \partial _\beta  (\sqrt g g^{\alpha \beta } \partial _\alpha  \psi ^a )
- 2\sqrt g \psi ^a  + 2\sqrt g g^{\alpha \beta } \omega _\alpha ^{[ab]} \partial _\beta  \psi ^b  = O(t^2 \psi )
\end{equation}                                                                                                                          
To solve this equation we start from its asymptotic form as $t\to 0$, treating the other terms as small 
perturbations. At this point it becomes very convenient to introduce, following Refs [17,18], the
Fourier transform
\begin{equation}
\phi ^a (t,s) = \phi ^a (t,s' + h) = \int\limits_{ - \infty }^\infty  {{{dp} \over {2\pi }}} e^{iph} 
\tilde \phi ^a (t,p),
\end{equation}
 with $\sigma=s-{h\over 2}=s'+{h\over 2}$, the point at which the area derivative is applied. The
relevant observation here is that one is interested in large values for the variable $p\sim{1\over h}$,
since the variable $h$ is integrated in the vicinity of zero,cf. Eq(4) in the text.

On the other hand, one can be convinced, by appealing to (A.19), that the values of $t$ which are involved in 
our analysis are $t\sim {1\over |p|}\sim h$. With these estimations (A.18) can be rewritten by retaining
only those terms that are relevant to the normal variation of the $\vec{g}$-function. To accomplish this task the 
coefficient functions must be expanded around the point $s'$. The general form of such an expansion can be read from                                                                                                                            
\begin{equation}
\begin{array}{cc}
F(s) = F(s') + (s - s')F'(s') + ... = F(s') + hF'(s') + ...\\
h\phi ^a (t,s) = \int\limits_{ - \infty }^\infty  {{{dp} \over {2\pi }}e^{iph} } h\tilde \phi ^a (t,p) 
= \int\limits_{ - \infty }^\infty  {{{dp} \over {2\pi }}e^{iph} } i\partial _p \tilde \phi ^a (t,p)\\
\end{array}
\end{equation}
Given the above, Eq. (A.18) reads, in Fourier space,                                                                                                                               
\begin{equation}
\hat L_4^{ab} (t,p)\tilde \phi ^b (t,p)=\hat L_2^{ab}(t,p)\tilde \phi ^b (t,p) + 
\hat L_1^{ab} (t,p)\tilde \phi ^b (t,p) + ...,
\end{equation}
where we have written                                                                                                                              
\begin{equation}
\begin{array}{cc}
 \hat L_4^{ab}  \equiv ({1 \over {t^2 }}\partial _t^2  - {2 \over t}\partial _t  - {{p^2 } \over {t^2 }})
\delta ^{ab},\quad \hat L_2^{ab}  \equiv \vec f^2 (\partial _t^2  + p^2 )\delta ^{ab}.\\
\hat L_1^{ab}  \equiv \left\{ {\left[ {2\vec f \cdot \vec f'i\partial _p  + 
{4 \over 3}t(\vec f \cdot \vec g)} \right](\partial _t^2  + p^2 ) + {4 \over 3}\vec f \cdot \vec g\partial _t  
- {3 \over 2}\vec f \cdot \vec f'ip + t\vec f \cdot \vec f'ip\partial _t } \right\}\delta ^{ab}  
+ r^{ab} ({1 \over t} - \partial _t )\\ 
\end{array}
\end{equation}
The subscripts labeling the operators in the above relation serve to signify their asymptotic behavion as $|p|\to\infty$:
\begin{equation}
\hat L_4^{ab} \tilde \phi ^b  \sim O(p^4 ),\quad\hat L_2^{ab} \tilde \phi ^b  
\sim O(p^2 ),\quad\hat L_1^{ab} \tilde \phi ^b  \sim O(p).
\end{equation}                                                                                                                               
The neglected terms in (A.21) are of order ${\cal O}(p)$ so that their contribution will be
four times weaker that the strongest one and thus irrelevant as far as the 
normal variation of the $\vec{g}$-function.

The solution of (A.21) can be written as                 
 \begin{equation}
\tilde \phi ^a (t,p) = \tilde \phi _{(0)}^a (t,p) + \int\limits_0^\infty  
{dt'} G_p (t,t')\left[ {\hat L_2^{ab} (t',p) + \hat L_1^{ab} (t',p)} \right]\tilde \phi ^a (t',p)
\end{equation}
Here $\tilde{\phi}^a_{(0)}$ is the solution of the homogeneous equation                                                                                                                               
\begin{equation}
\begin {array}{cc}
\hat L_4^{ab} (t,p)\tilde \phi ^b (t,p) = 0\\
\tilde \phi _{(0)}^a (t,p) = (1 + t\left| p \right|)e^{ - t\left| p \right|} \tilde \phi _{(0)}^a (p)\\
\end{array}
\end{equation}
                                                                                                                               
The Green's function
\begin{equation}
\hat L_4^{ab} (t,p)G_p (t,t') = \delta (t - t')
\end{equation}
can be easily found:
\begin{equation}
G_p (t,t') = {1 \over {2\left| p \right|^3 }}\phi _ -  (t'\left| p 
\right|)[\phi _ +  (t'\left| p \right|) - \phi _ -  (t'\left| p \right|)]\theta (t - t') + (t \leftrightarrow t'),
\end{equation}                                                                                                                               
with
\begin{equation}
\phi _ -  (x) = (1 + x)e^{ - x} {\rm   }{\rm ,    }\phi _ +  (x) = (1 - x)e^x.
\end{equation}
The solution of the integral equation (A.24) can be approached through an iterative procedure:
\begin{equation}
\tilde \phi ^a (t,p) = \tilde \phi _{(0)}^a (t,p) + \int\limits_0^\infty  {dt'} 
G_p (t,t')\left[ {\hat L_2^{ab} (t',p) + \hat L_1^{ab} (t',p)} \right]\tilde \phi _{(0)}^a (t',p) + 
negligible\,\,terms
\end{equation}                                                                                                                               

Expanding, now the result in a $t$ power series one can see that the neglected terms in the above equation are of 
order ${\cal O}(t^4)$ and thus irrelevant for our purposes. The symmetric part of the solution (A.29) 
is easily determined to be
\begin{equation}
\left[ {1 - {1 \over 2}\left| p \right|^2 t^2  - {1 \over 3}t^3 (\vec f^2 \left| p \right| + 
i\vec f \cdot \vec f'signp + \vec f \cdot \vec g)} \right]\tilde \phi _{(0)}^a (p) + O(t^4 ),
\end{equation}
while the contribution to the antisymmetric part is                                                                                                                               
\begin{equation}
\int\limits_0^\infty  {dt'} G_p (t,t')({1 \over {t'}} - \partial _{t'} )e^{ - \left| p \right|t'} 
(1 + \left| p \right|t')r^{ab} \tilde \phi ^a  =  - {1 \over 3}t^3 [\Gamma (0,2\left| p \right|t) + 
{{25} \over {12}}]r^{ab} \tilde \phi ^a  + O(t^4 ).
\end{equation}
The next step is to integrate the `annoying' incomplete gamma function:                                                                                                                               
\begin{equation}
\int\limits_{ - \infty }^\infty  {{{dp} \over {2\pi }}} e^{iph} \Gamma (0,2t\left| p \right|) 
= 2{\mathop{\rm Re}\nolimits} \mathop {\lim }\limits_{\varepsilon  \to 0} \int\limits_0^\infty  
{dp} e^{iph} \Gamma (\varepsilon ,2t\left| p \right|) 
= 2{\mathop{\rm Re}\nolimits} \mathop {\lim }\limits_{\varepsilon  \to 0} {t \over {2ih}}\Gamma (\varepsilon )
[1 - {1 \over {(1 + {{ih} \over {2t}})^\varepsilon  }}] = {1 \over t}+{\cal O}(h)
\end{equation}
and thus the ${\cal O}(t^3)$ antisymmetric contribution to the solution can be taken to be just
\begin{equation}
- {1 \over 3}t^3 {{25} \over {12}}r^{ab} \tilde \phi ^a. 
\end{equation} 
To obtain the final result one must take into account that normal variations do not preserve the static gauge
and, therefore, a redefinition of the $t$ variable is needed. Repeating the relevant calculation 
of Ref [16] we arrive at Eq. (23) of the text.                     
                                                            
\newpage

\vspace*{0.5cm}

\begin{center}
{\bf Acknowledgement}
\end{center}

\vspace{0.1cm}

The authors wish to acknowledge financial supports through the
research program ``Pythagoras'' (grant 016) and by the General Secretariat of Research and
Technology of the University of Athens.


\begin{thebibliography}{99}


\bibitem{1} A. Polyakov, Phys. Lett. B 82 (1979)247. 

\bibitem{2} G. 'tHooft, Nucl. Phys. B 72(1974) 461.

\bibitem{3} A. Polyakov, Nucl. Phys. B (Proc. Suppl.) 68(1998) 1.

\bibitem{4} K. G. Wilson, Phys. Rev. D(1974) 2445. 

\bibitem{5} J. Maldacena, Adv. Theor. Math. Phys. 2 (1998) 231.

\bibitem{6} S. Gubser, I. Klebanov and A. Polyakov, Phys. Lett. B 428(1998) 105.

\bibitem{7}E. Witten, Adv. Theor. Math. Phys. 2 (1998) 253.

\bibitem{8}J. Maldacena, Phys. Rev. Lett. 80(1998) 4859. 

\bibitem{9}N. Drukker, D. J. Gross and H. Ooguti, Phys. Rev. D 60(1999) 125006. 

\bibitem{10}S.-J. Rey and J. Yee, Eur. Phys. J. C 22(2001) 379. 

\bibitem{11}J. Polchinski and M. J. Strassler, Phys. Rev. Lett. 88(2002) 031601; J. High Energy Phys. 05(2003) 012.   

\bibitem{12}S. D. Avramis, A. I. Karanikas and C. N. Ktorides, Phys. Rev. D 66(2002) 045017.

\bibitem{13}A. M. Polyakov, {\it Gauge Fields and Strings}, Harwood Academic Publishers, Switzerland, 1987.

\bibitem{14}A. M. Polyakov, Nucl. Phys. Â 164 (1979)171.

\bibitem{15}A. Polyakov and V. Rychkov, Nucl. Phys. B 581(2000) 116.

\bibitem{16}A. M. Polyakov and V. S. Rychkov, Nucl. Phys. B 594(2001) 272. 

\bibitem{17}R. A. Brandt, A. Gocksch, M. A. Sato and F. Neri, Phys. Rev. D 26(1982) 3611. 

\bibitem{18}A. I. Karanikas and C. N. Ktorides, JHEP 11(1999) 033.

\bibitem{19}Yu. M. Makeenko and A. A. Migdal, Phys. Lett. B 97(1980) 253. 

\bibitem{20}A. A. Migdal, Phys. Rep. 102 (1983) 199; Yu. M. Makeenko, {\it Methods of Contemporary Gauge Theory}, 
Cambridge Monograph on Mathematical Physics(2002). 

\bibitem{21} H. G. Dosch and Yu. A. Simonov, Phys. Lett. B 205(1988) 339.

\bibitem{22}A. I. Karanikas , C. N. Ktorides and N. G. Stefanis, Eur. Phys. J. C 26(2003) 445.

\end{thebibliography}
\end{document}